\documentclass[pra,aps,twocolumn,groupedaddress,superscriptaddress,floatfix]{revtex4}
\usepackage{epsfig}
\usepackage{amsmath}
\newcommand{\be}{\begin{equation}}
\newcommand{\ee}{\end{equation}}
\newcommand{\bea}{\begin{eqnarray}}
\newcommand{\eea}{\end{eqnarray}}
\newcommand{\ket}[1]{\left|#1\right\rangle}

\newcommand{\bc}{\begin{center}}
\newcommand{\ec}{\end{center}}

\newcommand{\forget}[1]{}

\begin{document}

\preprint{}
\title{Exploiting the quantum Zeno effect to beat photon loss in linear
optical quantum information processors }
\author{Federico M. Spedalieri}
\email{Federico.Spedalieri@jpl.nasa.gov}
\author{Hwang Lee} 
\author{Marian Florescu}
\author{Kishore T. Kapale}
\author{Ulvi Yurtsever}
\affiliation{Quantum Computing Technologies Group, Jet Propulsion Laboratory,
California Institute of Technology, Mail Stop 126-347, 4800 Oak Grove Drive, 
Pasadena, California 91109-8099}
\author{Jonathan P. Dowling}
\affiliation{Quantum Computing Technologies Group, Jet Propulsion Laboratory,
California Institute of Technology, Mail Stop 126-347, 4800 Oak Grove Drive, 
Pasadena, California 91109-8099}
\affiliation{Department of Physics and Astronomy, Louisiana State University,
Baton Rouge, Louisiana 70803-4001}
\date{\today}

\begin{abstract}
We devise a new technique to enhance transmission of quantum 
information through linear optical quantum information processors. 
The idea is based on applying the
Quantum Zeno effect to the process of photon absorption. By 
frequently monitoring the presence of the photon through a QND (quantum 
non-demolition) measurement the absorption is suppressed. Quantum information
is encoded in the polarization degrees of freedom and is therefore
not affected by the measurement. Some implementations of the QND measurement
are proposed.

\end{abstract}
\maketitle

\section{Introduction}

Photons are the information carriers in linear optical quantum 
information processing devices~\cite{knill2001a,
pittman2001a}. Information
is encoded in the state of the photon (e.g., photon number or polarization),
and the processing is accomplished by sending the photon through a system
of linear optical elements and using photodetectors to measure the outcome.
The photons are usually routed through optical fibers or waveguides, 
that can transmit
them from one part of the device to another and act as delay lines
or even as primitive memories, by letting the photon go around a loop of fiber.
This useful property of a fiber as a transmission line has its origin
in the interaction between the photon and the atoms in the fiber. However,
this interaction has the undesirable side effect of photon absorption,
which introduces errors in the quantum information processing.
Thus it has been an active area of research
to develop new approaches to suppress photon absorption in optical 
processors~\cite{wu2003a}.

Since we cannot amplify a quantum signal, our efforts should be aimed
at preventing its degradation due to the loss of photons caused
by the interaction between the fiber and the electromagnetic field.
For this purpose, we propose to use the well-known Quantum Zeno
effect~\cite{misra1977a}, which has already been applied to quantum 
information processing~\cite{franson2004a}. 
To understand the idea we note that 
this loss of photons can be viewed as the evolution of a quantum state
in which the initial state contains one photon and, 
due to the interaction between the photon and the fiber, is transformed 
into a state in which the photon has been absorbed. The Quantum 
Zeno effect tells us that by frequently 
monitoring the presence of the photon we can inhibit this 
evolution, and hence prevent the photon from being absorbed by the fiber.
It may seem rather strange that we propose to preserve the quantum
information encoded in the system by measuring it, since measurement
usually destroy quantum coherence. The key idea is to encode the
information in the polarization degrees of freedom of the photon, while
performing a quantum non-demolition measurement only on the photon
number. In this way no information about the polarization state
is acquired, but frequently registering the photon presence
prevents it from being absorbed.

\section{The model}
 
To simplify our discussion, we will couch our argument in terms of 
single photon loss in telecom fibers. However, the scheme 
applies to loss in any
type of linear optical quantum information processor.

To start our analysis we first need to identify the physical
mechanisms that produce photon loss on a fiber. At the usual telecom
wavelength of $1.55 \mu m$, there are two main loss mechanisms
whose deleterious effect on the photon transport  have approximately the 
same magnitude:
(i) the Rayleigh scattering of photons on the inhomogeneities of
the refractive index along the fiber; (ii) the absorption of photons 
due to their interaction with the vibrational excitations 
(phonons)~\cite{kasap2002a}. 
In principle,
the effect of Rayleigh scattering can be lowered by improving 
construction to minimize the inhomogeneities in the fiber~\cite{saito2003a}.
On the
other hand, the interaction with phonons will always be present,
and hence it is useful to design a procedure to reduce its effects.

Our goal then is to analyze the interaction between the signal photon and 
the phonons in the fiber, and show that we can prevent the loss of the
photon due to this interaction. For this purpose we consider the
following Hamiltonian 
\begin{equation}
H = H_{signal} + H_{bath} + H_{int},
\end{equation}
where $H_{signal} = \sum \hbar \omega_{\mathbf{k}} a_{\mathbf{k}}^\dagger 
a_{\mathbf{k}}$ is the Hamiltonian 
of the propagating signal photon, $H_{bath} = \sum \hbar \omega_i r_i^\dagger
r_i$ is the Hamiltonian of the phonon bath, and 
$H_{int} = \sum_{i \mathbf{k}} (g_{i \mathbf{k}} a_{\mathbf{k}} r_i^\dagger +
g_{i \mathbf{k}}^\ast a^\dagger_\mathbf{k} r_i)$,
is the interaction
Hamiltonian. The precise form of $H_{int}$ is not essential to understand
the idea behind our scheme, although it is important in determining
the regime of validity  of this proposal.    

The signal will consist of a photon pulse of the form
$|\psi_{signal}\rangle = \sum_\mathbf{k} f(\mathbf{k}) |1_\mathbf{k}
\rangle (\alpha |H\rangle +\beta |V\rangle)$,
where $|H\rangle$ and $ |V\rangle$ represent horizontal and vertical
polarizations respectively. The quantum information carried by the signal
is encoded in the values of the coefficients $\alpha$ and $\beta$. 
The function $f(\mathbf{k})$ is chosen such that the pulse has a finite 
spatial extent and the mean number of photons in the pulse is 1. 

The initial state of the photon and the fiber can be written as
$|\Psi \rangle_{t=0} = |\psi_{signal}\rangle |\phi_{bath}\rangle$,
where $|\phi_{bath}\rangle$ is the initial state of the phonons 
in the fiber. To simplify the notation, let us assume that initially
the we have no phonons in the fiber, so we have $|\phi_{bath}\rangle =
|000\cdots\rangle$. As the pulse propagates, the 
interaction part of the Hamiltonian will entangle the photons with the
phonon in the fiber. At some later time $t$, the state of the system
will have the form
\begin{eqnarray}
\label{psit}
|\Psi (t)\rangle & = &  \left(\sum_\mathbf{k} f(\mathbf{k},t) |1_\mathbf{k}
\rangle (\alpha |H\rangle +\beta |V\rangle)\right)|000\cdots\rangle + \nonumber \\
& & + \sum_j c_{0j}(t) |0\rangle_{photon} |0\cdots \underbrace{1}_{j^{th} 
mode} 0 \cdots \rangle.
\end{eqnarray}
The first sum represents the free propagation of the photon, while the
second sum shows that the photon can be absorbed only by creating 
a phonon in one of the modes of the fiber. To find the time
dependent functions $f(\mathbf{k},t)$ and $ c_{0j}(t)$ we can insert
(\ref{psit}) into the Schr\"odinger equation and solve the resulting
differential equations. However, this will not be necessary for our 
particular purpose.

Our goal is to use the Quantum Zeno effect to suppress the absorption
of the photon due to its interaction with the phonons in the fiber.
This is accomplished by performing frequent quantum non-demolition
(QND) measurements of the photon number of the pulse. If we measure
the photon number at time $t=\tau$ and the outcome is 1, then the
state of the system after the measurement is given by
\begin{equation}
\label{psitau}
|\Psi (\tau)\rangle =   \left(\sum_\mathbf{k} f(\mathbf{k},\tau) |1_\mathbf{k}
\rangle (\alpha |H\rangle +\beta |V\rangle)\right)|000\cdots\rangle. 
\end{equation}
Since the photon number QND
measurement does not affect the polarization degrees of freedom,
\emph{the quantum information carried by them is not affected by
the measurement}. This is the key idea of our proposal. The state 
in Eq. (\ref{psitau}) is the new initial condition
to be propagated for another interval $\tau$ until the next QND
measuremnet. Note that this initial condition might be different from
the one at $t=0$, since the functions $f(\mathbf{k},\tau)$ might differ 
from the functions $f(\mathbf{k})|_{t=0}$. However this is not a problem,
because the quantum information carried by the pulse is independent
of its form.

The Quantum Zeno effect tells us that these frequent QND 
photon number measurements can prevent photon absorption. To see 
how this works, let $P_s (t)$ be the survival probability 
of the photon pulse at time $t$ as it propagates through
the fiber. From Eq. (\ref{psit}) we have that 
$P_s (t) = \sum_{\mathbf{k}} |f(\mathbf{k},t)|^2$.
Due to the unitarity of quantum evolution, is a well-known fact
that this survival probability must follow a quadratic decay
for small values of $t$. If the system is interacting with a 
continuum of modes,
it is also well-known that this decay becomes exponential for
longer times. The time scale at which the transition occurs between
these two regimes strongly depends on the details of the interaction
and the physical properties of the interacting systems. We will come back
to this point later. For now let us analyze the effects of frequent
measurements during the quadratic decay. Let $\tau$ be the time interval
between measurements. The survival probability at time $\tau$
will be given by $P_s (\tau) \simeq 1 - (\gamma\tau)^2$, 
for some constant $\gamma$. With probability $P_s (\tau)$ the outcome
of the photon munber QND measurement is 1, and we know with certainty that 
the pulse has exactly 1 photon after the measurement. This pulse
evolves again for a time $\tau$, and the probability of it containing
exactly 1 photon is again given by $P_s (\tau)$. After performing
$N$ QND masurements separated by intervals $\tau$, the probability
of survival at time $T= N \tau$ is given by
\begin{eqnarray}
P_s (T) & = & \left( 1- (\gamma\tau)^2\right)^N \nonumber \\
        & = & \left( 1- \frac{T \tau\gamma^2}{N}\right)^N \nonumber \\
        & \simeq & e^{- (\gamma^2\tau) T},
\end{eqnarray}
where we considered the large $N$ limit on the last line.
We see that the decay is still exponential, \emph{but the decay rate
depends of the parameter $\tau$}. By choosing smaller values of
$\tau$ (i.e., increasing the frequency of the measurements), we can
suppress the decay rate and hence diminish the probability of the
photon being absorbed.   

To apply this idea to the problem of photon loss,
we consider the usual telecom fiber and insert a QND 
measurement devices at regular intervals. Since the succesful
application of the Quantum Zeno effect requires that 
successive measurements be made while the decay of the survival
probability is quadratic, and this occurs for times shorter than a
certain value $T_q$, the QND devices must be placed such that
the distance between them is less than $v_f T_q$, where $v_f$ is the
velocity of the pulse in the fiber. 

The time scale given by $T_q$ is an important parameter in our proposal
since it determines how close together the QND devices must be inserted
along the fiber. A calculation of this value from first principles appears
to be a daunting task, since it will depend crucially on the details
of the photon-phonon interaction, and in particular on the exact form
of the density of phononic modes on the fiber.
 A simplified calculation of this time scale
is highly dependent on the approximations used to model the density
of states, giving answers that differ by several orders of 
magnitude~\cite{wu2003a}. This suggests that this type of 
calculation would not be reliable unless the density of modes is
known with great detail.  However this time scale should not be difficult
to measure. To that end we propose the following simple experiment:
measure the single photon loss for different lengths of fiber, starting with a 
length for which exponential decay of the output signal has been established,
and repeat the experiment for shorter lengths until a departure from the
exponential decay is observed. That will happen about the time scale
given by $T_q$, for which the decay becomes quadratic.  

It is interesting to note that the same idea we used to enhance
the transmission of quantum information through a fiber or other
linear optical quantum information processor, can be
used to design an improved memory device for optical quantum computation.
Instead of inserting many QND devices along a transmission line, we can
just insert only one device on a loop of fiber. Again by encoding
the quantum information on the polarization degrees of freedom of the
signal pulse and performing a QND measurement of the photon number, we 
can suppress the absorption of the pulse by the fiber, and hence increase 
the time we can store the quantum information in the fiber. This alternative
use
could be viable even if the transmission application turns out not to
be practical because it requires too many QND devices.

\section{QND devices}

We will now present a few ideas for the implementation of the QND
device required by our proposal. First, let us note that we only need
a QND photon number measuring device that can distinguish between
1 and 0 photons. We do not need it to be able to discriminate for higher
photon number, since our signal pulse will be constructed such that it
contains only one photon. Then, a simple CNOT gate acting on the 
$\{|0\rangle,|1\rangle\}$ subspace and one ancilla mode are 
sufficient to implement
the QND measurement. If we prepare the ancilla in the $|0\rangle$ state,
and use it as the target of a CNOT gate controlled by the presence or
absence of a photon in the fiber, by measuring the ancilla we can determine
whether the gate has been applied or not, and hence whether there was
a photon in the fiber or not. 
We can even apply the KLM 
~\cite{knill2001a} scheme that uses 
linear optical elements to
implement the CNOT gate. This implementation of the gate is 
non-deterministic but,
its probability of success can be boosted arbitrarily close to unity by 
adding more and more ancillas.

We now proceed to give possible implementations for the QND device in quantum optical systems. 
Strong nonlinearities are essential to achieve a non-destructive measurement of the photon number state. 
Such strong nonlinearities can be attained through either atomic coherence effects, for example, 
electromagnetically induced transparency (EIT) or strong atom-field interaction obtained in the cavity-QED
schemes. 

An interesting proposal for implementation of the QND device exists that can be adopted for the present scheme. 
Munro {\it et al.}~\cite{munro2003a} have proposed a high-efficiency QND single
photon number resolving detector based on cross-Kerr nonlinearity obtainable through EIT. 
To illustrate its mechanism,
we note that the QND Hamiltonian can be written as
$
H_{\rm QND} = \hbar \chi\,s^{\dagger}s\,p^{\dagger} p\,, 
$
where $s$ and $p$ are the ladder operators for the signal and probe fields respectively.
The input signal in a fock state $\ket{n_s}$ can be detected non-destructively through the phase
acquired by a coherent probe field $\ket{\alpha_p}$:
\begin{equation}
\label{Eq:phase}
|\Psi(t)_{\rm out}\rangle= {\rm e}^{i \chi t s^{\dagger}\, s\, p^{\dagger}
  p}|n_s \rangle |\alpha_p\rangle
  =|n_s \rangle|\alpha_p \, {\rm e}^{i n_s \chi t}\rangle\,.
\end{equation}
Thus, a momentum quadrature measurement on the probe field through an efficient homodyne device ascertains the presence of the signal photon non-destructively. 

\begin{figure}[ht]
\centerline{\includegraphics[scale=0.24]{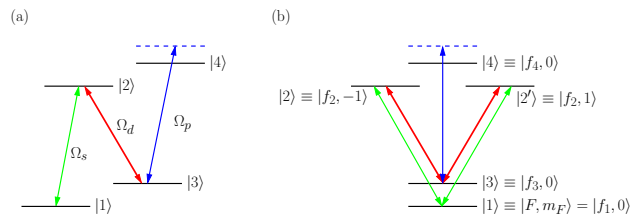}}
\caption{\label{Fig:LevelSchemes} EIT based QND device (level schemes): (a) A general 4-level
scheme: QND measurement of a green signal photon ($\Omega_s$) is achieved through a phase quadrature 
measurement of the highly-detuned blue probe field ($\Omega_p$). The strong red driving field ($\Omega_d$) 
facilitates EIT.  (b) Level scheme for the polarization-insensitive QND device: 
The levels are denoted by a pair of hyperfine quantum numbers as $\ket{F,m_F}$. The color code for various fields is the same as in (a). Probe and drive fields are linearily
polarized to facilitate extraction of the photon number of the signal photon immaterial of its linear or circular
polarization state.}
\end{figure}

This Kerr-nonlinearity based QND device can be implemented through a 4-level EIT scheme shown in
Fig.~\ref{Fig:LevelSchemes}(a). The atomic medium starts in the initial state
$\ket{1}$ and remains in that state after the interaction with the signal except of an overall phase
acquired by the probe according to Eq.~(\ref{Eq:phase}). A momentum-quadrature measurement is performed on the 
probe field to reveal the presence of the signal photon non-dectructvely. 
Munro {\it et al.}~\cite{munro2003a} carry out a detailed analysis of this four level scheme and obtain various conditions 
necessary for efficient QND measurement on the signal photon. 

It is, however, important to note that the Munro scheme, when applied to an actual atomic system,  
would depend on and affect the polarization state of the signal photon. For the success of our
proposal the QND device should be insensitive to the polarization of the incoming photon. Moreover, the
measurement process should neither change this polarization state nor leave its imprint on the atom or the probe. 

We note that radiative atomic transitions couple to a specific polarization states of light and devise an 
implementation of a polarization-insensitive QND device using hyperfine sublevels of a certain atom as shown in
Fig.~\ref{Fig:LevelSchemes}(b). It can be easily seen that model in Fig.~\ref{Fig:LevelSchemes}(b) has 
two level schemes of Fig.~\ref{Fig:LevelSchemes}(a) (namely $\ket{1}-\{\ket{2},\ket{2^\prime}\}-\ket{3}-\ket{4}$) 
embeded in it. We require the drive and probe fields to be linearily polarized. Thus, the drive field couples to 
both $\ket{2}-\ket{3}$ and $\ket{2^\prime}-\ket{3}$ transitions according to the hyperfine-transition selection 
rules. This is possible because linearily polarized field can be represented as a superposition of equal-magnitude
left- and right-circularly polarized components.  Similarly, through appropriate choice of hyperfine
quantum numbers, the linearily polarized probe field can be set to couple a single $\ket{3}-\ket{4}$
transition. It can be easily seen that the probe field would not carry any information about the polarization of 
the signal photon as the oscillator strengths of $\ket{1}-\ket{2}$ and $\ket{1}-\ket{2^\prime}$ transitions 
are the same. Thus, immaterial of the polarization state of the incoming 
photon (linearily or circularly polarized) the level scheme of Fig.~\ref{Fig:LevelSchemes}(b) would respond to it 
in the same way and a quadrature measurement on the probe would neither disturb nor reveal this state. 

It can be noted that the level scheme used in  Fig.~\ref{Fig:LevelSchemes}(b) is fairly general, 
as the values for the quantum number $F$,  namely $f_1, f_2, f_3,$ and $f_4$, 
are unspecified and only the specific $m_F$ values are assumed. 
One can come up with various examples of atoms that have level-scheme matching the one proposed here. 
A detailed analysis of this scheme will be carried out elsewhere~\cite{kapale2004a}. It is useful to note that such a
polarization-insensitive QND device could be realized in variety of systems including cold atomic clouds, single
alkali atom trapped in a high-Q cavity and rare earth doped telecommunication fiber. 

Furthermore, we note that the quantum phase gate ($\ket{c,t}\rightarrow\exp(i \phi \delta_{c,1}
\delta_{t,1})\ket{c,t}$ with a phase shift ($\phi=\pi$) and single qubit rotations (on $\ket{t}$) could be
combined to perform a CNOT operation $\ket{c,t}\rightarrow\ket{c,c\oplus t}$, where $\ket{c}$ and $\ket{t}$ are
the control and target qubits respectively. This opens up another powerful possibility for the
implementation of the QND device, through the CNOT gate, based on well developed tools of cavity-QED. 
There are both experimental~\cite{rauschenbeutel1999a} and theoretical~\cite{zubairy2003a} proposals for realizing a quantum phase gate in
cavity-QED systems for photonic qubits. One can easily adopt these techniques by coupling the signal photon to 
the cavity to perform the role of the control qubit. Once again the main requirement would be to make these 
schemes insensitive to the polarization of the signal photon. 

The cavity-QED techniques can be easily extended to the regime of fiber optics by replacing the bulky
superconducting cavities by either Fabry-Perot cavities, which are much easier to integrated into the
telecommunication fiber, or the microresonators, which can be easily coupled to the fiber through the evanescent 
field of the signal photon. To note, Kimble's group at caltech 
has done a lot of work on extending cavity-QED to both the Fabry-Perot cavities and microresonators~\cite{buckthesis}.

In this letter we have presented a new scheme to enhance the transmission
of quantum information through linear optical quantum information
processing systems by taking advantage
of the Quantum Zeno effect. We propose encoding the quantum information
in the polarization degrees of freedom of the photon. To overcome the 
losses due to absorption of the photon by the fiber, we propose to
frequently monitor the presence of the photon by performing a QND measurement 
of the photon number. Due to the Quantum Zeno effect, this procedure
can supress the absorption if the measurements are performed frequently
enough. The quantum information carried by the photon is not disturbed
since the polarization state is not affected by the QND photon number
measurement. We also presented some possible implementations of this 
QND measurement, and proposed an experimental method to determine how
often the measurements should be performed. This scheme could also
be used as a memory, by preventing a photon stored in a loop
of fiber from being absorbed, hence preserving the quantum 
information it carries for a longer time.

This research has been carried out at the Jet Propulsion Laboratory under 
a contract with the National Aeronautics and Space Administration (NASA). 
We would like to acknowledge support  from the Advanced Research and
Development Activity, the National Security Agency, The Defense Advanced
Research Projects Agency, the Office of Naval Research, the National 
Reconnaissance Office, and the NASA Intelligent Systems Program. In addition,
FMS and KTK acknowledge support from the National Research Council and 
NASA, Code Y, and MF
acknowledges support from the National Research Council and NASA, Code S.

\bibliographystyle{prsty}
\bibliography{Cav}

\end{document}